# Denser glasses relax faster: a competition between rejuvenation and aging during *in-situ* high pressure compression at the atomic scale

A. Cornet[a,b]*, G. Garbarino[b], F. Zontone[b], Y. Chushkin[b], J. Jacobs[b], E. Pineda[c], T. Deschamps[a], S. Li[a], A. Ronca[a,b], J. Shen[a,b], G. Morard[d], N. Neuber[e], M. Frey[e], R. Busch[e], I. Gallino[e], M. Mezouar[b], G. Vaughan[b], and B. Ruta[a,b]*

[a]*Institute of Light and Matter, UMR5306 Université Lyon 1-CNRS, Université de Lyon, F-69622 Villeurbanne, France*

[b]*European Synchrotron Radiation Facility, 71 avenue des Martyrs, CS 40220, Grenoble 38043, France*

[c]*Department of Physics, Institute of Energy Technologies, Universitat Politècnica de Catalunya—BarcelonaTech, 08019 Barcelona, Spain*

[d]*Université Grenoble Alpes, Université Savoie Mont Blanc, CNRS, IRD, Université Gustave Eiffel, ISTerre, 38000 Grenoble, France*

[e]*Chair of Metallic Materials, Saarland University, Campus C6.3, 66123 Saarbrücken, Germany*

*corresponding authors
antoine.cornet@univ-lyon1.fr
beatrice.ruta@univ-lyon1.fr

**Abstract**

A fascinating feature of metallic glasses is their ability to explore different configurations under mechanical deformations. This effect is usually observed through macroscopic observables, while little is known on the consequence of the deformation at atomic level. Using the new generation of synchrotrons, we probe the atomic motion and structure in a metallic glass under hydrostatic compression, from the onset of the perturbation up to a severely-compressed state. While the structure indicates reversible densification under compression, the dynamic is dramatically accelerated and exhibits a hysteresis with two regimes. At low pressures, the atomic motion is heterogeneous with avalanche-like rearrangements suggesting rejuvenation, while under further compression, aging leads to a super-diffusive dynamics triggered by internal stresses inherent to the glass. These results highlight the complexity of the atomic motion in non-ergodic systems and support a theory recently developed to describe the surprising rejuvenation and strain hardening of metallic glasses under compression.

**Introduction**

Every glass has its own story, which is encoded in the evolution of its properties. Once a glass is formed by rapidly cooling from the melt, its final state depends on the applied temperature protocol and spontaneously evolves with time[1]. Fast cooling rates create glasses trapped in more energetically unstable configurations with larger structural disorder, lower elastic moduli and larger frozen-in free volume than slow cooling protocols [2]. Upon successive annealing, the glass ages and relaxes towards energetically more stable minima in the potential energy landscape (PEL), exploring continuously different configurations. This process, called physical aging, is particularly strong in metallic glasses (MGs), and modifies the mechanical, structural and thermal properties of the material [3,4]. In stark contrast, fast thermal cycling or mechanical deformation can rejuvenate the system, driving the glass into energetically less favoured configurations with increased plasticity [5–12]. In some cases, the rejuvenated MG would be equivalent to quenched glasses theoretically obtainable with cooling rates much faster than those reachable in a laboratory [6]. In the presence of almost hydrostatic compression, this rejuvenation leads to the suppression of shear banding and the inhibition of catastrophic mechanical failures, making deformed MGs appealing for technological applications [7]. Diffraction studies suggest a broadening of the interatomic distances in severely deformed MGs, which is in opposite to the well-known increase of structural order during physical aging [13]. Changes in correlation lengths at medium range order (MRO) have been also reported in pre-deformed Pd-based MGs and are accompanied by an acceleration of the microscopic relaxation dynamics possibly due to an increase in free volume [14,15].

The majority of studies deal with ex-situ compressed glasses, while little is known on the microscopic physical mechanisms occurring during the compression, owing to the experimental difficulty of in-situ experiments under high pressures. Theoretical works ascribe the pressure-induced rejuvenation and strain hardening of MGs to the creation of an additional local minimum in the PEL associated to rearrangements of the energy for cage dynamics [16,17]. This process would lead to the occurrence of two distinguishable dynamical regimes under pressure, whose existence has not been experimentally observed so far. By combining in-situ high pressure, high energy X-Ray Photon Correlation Spectroscopy (XPCS) and high energy X-ray Diffraction (XRD) in a 4$^{th}$ generation synchrotron source, we here provide the experimental evidence of how the atomic dynamics evolve under the application of pressure in a $Pt_{42.5}Cu_{27}Ni_{9.5}P_{21}$ MG, unveiling a complex, non-monotonous behaviour which is in agreement with recent theoretical works.

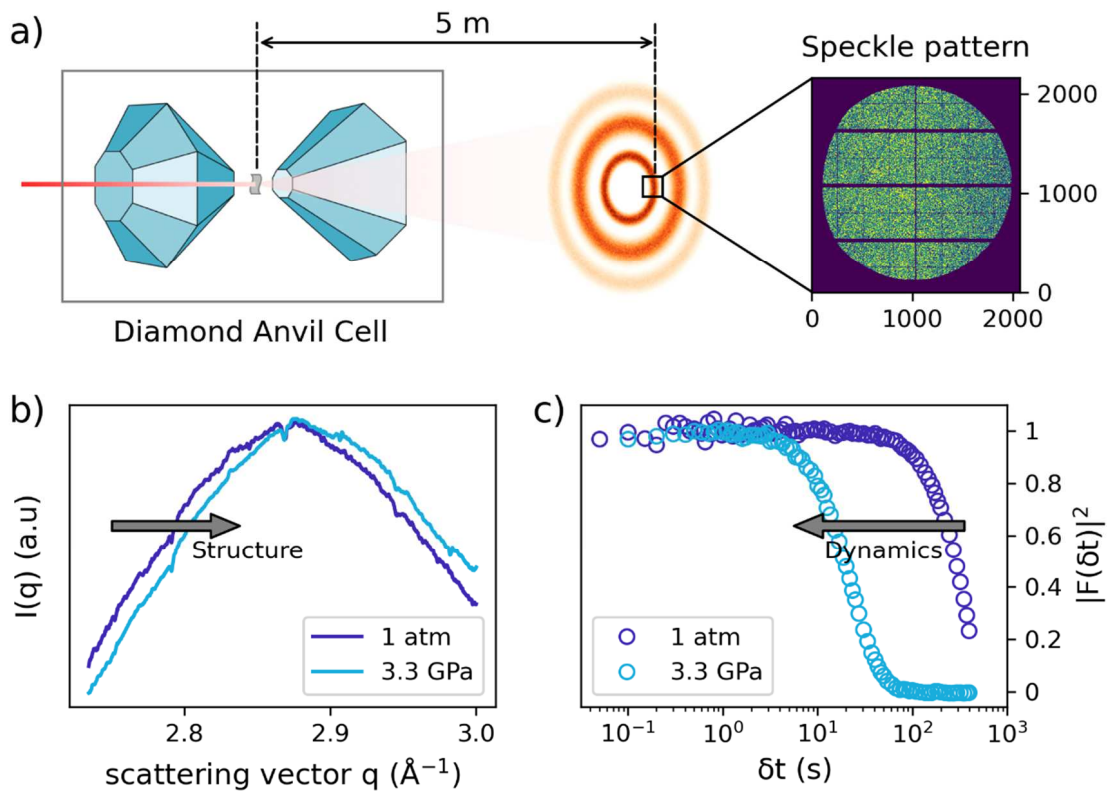

**Fig. 1 | Dynamical rejuvenation under densification at room temperature.** a) Sketch of an XPCS experiment showing the sample within the diamond anvil cell, the diffracted intensity corresponding to the structure factor, the portion of reciprocal space probed by the detector and a typical speckle pattern. b) top of FSDP covered during the XPCS experiments (the intensity

integrated across the detector area) measured at atmospheric pressure and at 3.3 GPa. Glitches in the I(Q) comes from the detector. c) Corresponding Intermediate Scattering Functions showing the acceleration of the dynamics with pressure.

**Results**

So far, the relatively low flux of high-energy coherent x-rays in 3$^{rd}$ generation synchrotrons, limited the use of XPCS in bulky sample environments, including diamond anvil cells (DACs) and other high-pressure apparatus. The current development of 4$^{th}$ generation synchrotron sources, such as the upgraded ESRF synchrotron (France), provides a monochromatic high flux ($10^{12}$ photon/s) of coherent x-rays at energies as high as 21 keV [18] with an unprecedented high quality, allowing for pressure dependent studies. A schematic view of the XPCS experiments is shown in Fig. 1a) and S1. The scattered speckle pattern is collected in a wide angle geometry covering the maximum of the first sharp diffraction peak (FSDP) of the glass which is at about $q_1$=2.87 Å$^{-1}$ for our as-cast $Pt_{42.5}Cu_{27}Ni_{9.5}P_{21}$ MG at ambient temperature and atmospheric pressure (Fig. 1b). By increasing pressure, the maximum of the FSDP shifts toward high scattering vectors, as shown in Fig. 1b) for 3.3 GPa. As the FSDP originates from the medium range order, in the absence of important structural rearrangements its position can be related to the macroscopic density of the glass [19]. The agreement between the thermal expansion coefficients of 3.85x10$^{-5}$ K$^{-1}$ obtained from us with high energy XRD (Supplementary Fig. S1), and the 3.95x10$^{-5}$ K$^{-1}$ value reported in literature dilatometry data [20] supports the validity of the density-MRO relation for the $Pt_{42.5}Cu_{27}Ni_{9.5}P_{21}$. Therefore, the continuous shift toward high scattering vectors, q, in Fig. 1b) reflects the monotonic rise in the glass density as the pressure increases.

The evolution of the internal dynamics of the glass accompanying the density change can be described by the pressure dependence of the intermediate scattering function (ISF), F($\delta$t), which monitors the temporal decay of the electron density fluctuations at the probed *q* and pressure. As shown in Fig. 1c), a pressure increase from 1 atm to 3.3 GPa results in a dramatic shift of more than one order of magnitude in the ISF toward smaller $\delta$t, which implies a pressure-induced acceleration of the atomic dynamic of the same magnitude. Fitting the data with the Kohlrausch-Williams-Watts (KWW) phenomenological model $|F(\delta t)|^2 = e^{-2(\delta t/\tau)^\beta}$ with τ the relaxation time and β the shape exponent, we find a relaxation time τ = 571s and τ = 38s for 1 atm and 3.3 GPa respectively at 300K. This pressure-induced acceleration of the dynamics by a factor 15 suggests a rejuvenation of the glass under *in-situ* hydrostatic compression and is larger than that observed in *ex-situ* deformed Pd-based MGs (factor 3.2 at 300K) [14] and around a single isolated shear band (factor 3.3 at 300K) in a $Zr_{65}Cu_{25}Al_{10}$ glass [21].

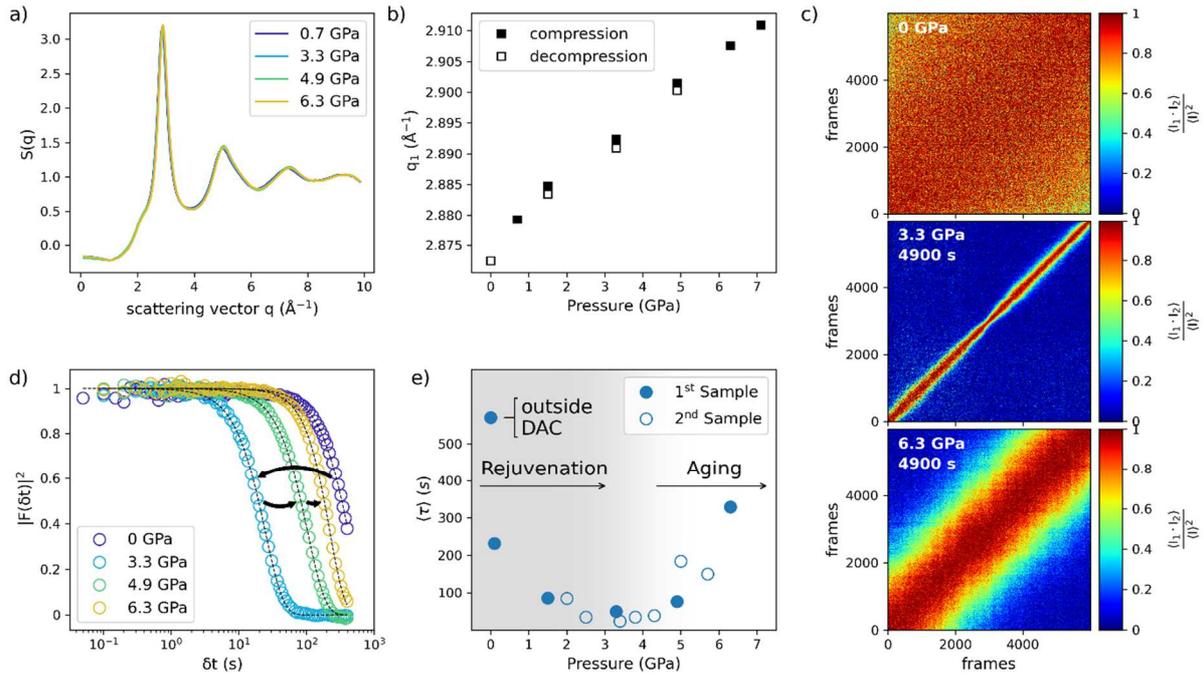

**Fig. 2 | Pressure dependence of structure and dynamics.** a) Static structure factor measured with high energy XRD under compression. b) corresponding maximum of the FSDP during both compression and decompression. c) TTCFs of selected scans acquired at 0 (1 atm), 3.3 and 6.3 GPa for similar elapsed times after the pressure change. d) Selected ISFs showing the transition from rejuvenation to relaxation with increasing pressure. Black dotted lines correspond to KWW fits to the data. e) Averaged relaxation time during compression (full symbols). Data of a second as-cast sample measured during a different XPCS experiment are reported as well to confirm the reproducibility of the results (empty symbols).

The overall evolution of the structure and dynamics during HP compression is reported in Fig. 2. The static structure factor, S(Q), varies only slightly with pressure, and exhibits a (4.91x10$^{-3}$ Å$^{-1}$/GPa) linear shift of $q_1$ with pressure up to 7 GPa, which is completely reversible with no hysteresis within the uncertainty of our measurement (Fig. 2a) and b)). In contrast, the collective atomic dynamics exhibits a complex evolution during the compression stage as shown by selected two-times correlation functions (TTCFs, Fig. 2c) and ISFs (Fig. 2d) measured after similar elapsed times from the pressure change at the different nominal pressures. The TTCF is a time-resolved representation of the ISFs, where the width of the high correlation contour is proportional to τ, the characteristic time of the rearrangements at the microscopic length scale.

At atmospheric pressure, high correlation values remain for most of the scan, which indicate almost arrested dynamics in the 600s total acquisition time. This is the classical picture of a glass well below the glass transition temperature, where large-scale dynamics are frozen and only slow local atomic rearrangements occur. The dramatic pressure-induced acceleration at 3.3 GPa corresponds to a sharp and narrow high-correlation contour and a fast decay of the ISF. As pressure increases from 3.3 GPa to 6.3 GPa, a non-monotonous evolution of the dynamics occurs, as evidenced by the larger width of the high correlation contours in the TTCFs and the shifts of the ISF to slower dynamics at high pressure.

To better clarify the nature of the atomic motion under hydrostatic compression, Fig. 2e) show relaxation times averaged over different scans acquired over a period of 3h at each single pressure, covering thus both the early stage deformation and the severely-compressed state. Two distinct dynamical regimes can be identified. An acceleration of the particle dynamics up to 3.3 GPa, followed by a progressive slow down at larger pressure values, suggesting the existence of a rejuvenation and a relaxation regime at low and high pressure, respectively. These results have been confirmed by repeating the experimental protocol in a different XPCS experiment on a second sample (see Methods for further details). Interestingly, the pressure-induced acceleration of the dynamics is visible even at the lowest pressure of 0.1 GPa, which corresponds to the preloading of the cell, where hardly any structural change is visible from XRD, showing the great sensitivity of the dynamics with respect to pressure. Artefacts related to the cell assembly, including the PTM, have been excluded as they give rise only to a static background, free of any dynamical contribution (Fig. S8). It is interesting to note that although this acceleration of a factor 2.5 in a small pressure interval is significant, it remains small when compared to the pressure-induced shifts of the structural relaxation times reported in softer molecular liquid glass-former [22,23].

To visualize how the dynamic varies with time during isobars in both the rejuvenation and relaxation regimes, Fig. 3) reports TTCFs measured at the extremum pressures of 3.3 and 6.3 GPa as a function of the elapsed time after pressure change. At low pressure, rejuvenation leads to heterogeneous dynamics with relaxation times fluctuating around an average constant value, as evidenced by the variation on the thickness of the red contour in the TTCFs at 3.3 GPa. We rule out the presence of possible artefacts, such as fluctuations in the incident flux and potential sample movement (Fig. S2 and S3). At about 6800s and 7100s at 3.3 GPa, complete decorrelation happens over one pixel in the TTCF, which is evidence for massive atomic rearrangements with a time scale lower than our acquisition time of 0.1s, while a steady acceleration of the dynamics is visible after 11600 s. We note that this heterogeneous dynamical regime does not stabilize over time during our experiment, as fluctuations are still visible on the TTCF after 12000s in the severely compressed state.

In sharp contrast to the heterogeneous, rejuvenation regime, the TTCFs show smoothly and continuously slowing down dynamics at 6.3 GPa, with decorrelation times growing with the time elapsed since pressure change. The transition from the low-pressure heterogeneous but constant dynamics regime to the homogeneous, high-pressure aging regime is not sharp but continuous, as visible from the evolution of $\langle \tau \rangle$ in Fig. 2e) and the TTCFs at 4.9 GPa which shows this intermediate regime, where both physical aging and fast massive atomic rearrangements are observed (Fig. S3). The existence of the two dynamical regimes is confirmed also by the reproducibility of the results in a second experiment (Fig. S4). The last row of the Fig. 3) corresponds to the TTCFs acquired at 3.3 GPa during decompression. It is highly similar to the aging regime visible at 6.3 GPa in compression, and does not match the heterogeneous dynamics observed at the corresponding pressure in compression. The pressure evolution of the dynamics is therefore not fully reversible, and exhibits a hysteresis evolution with a slow-down of the atomic motion of even a factor 10 during decompression (Fig. S5), in contrast with the apparently elastic behaviour of the structure under deformation (Fig. 2b).

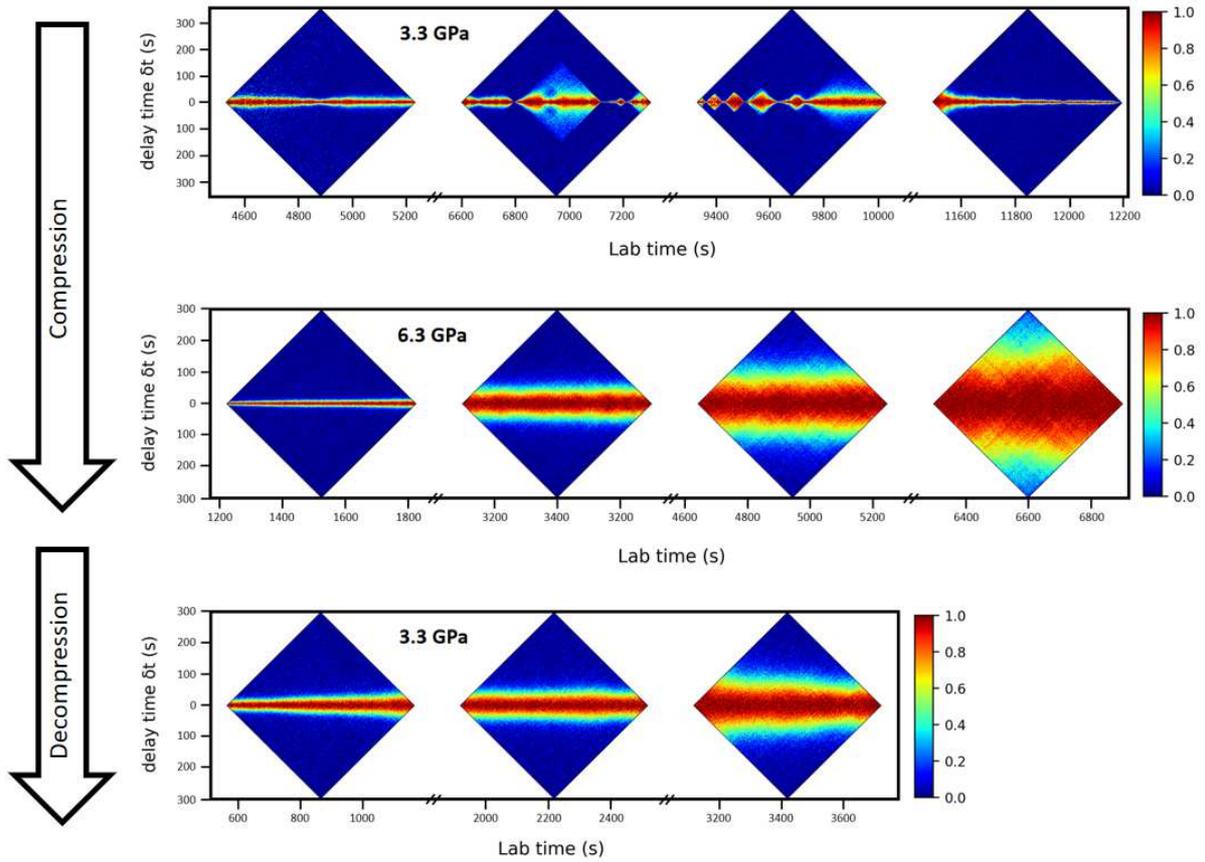

**Fig. 3 | Temporal evolution of the atomic motion during isobars.** TTCFs from scans acquired during compression (3.3 and 6.3 GPa) and during decompression (3.3 GPa), showing heterogeneous dynamics and physical aging at low and high pressure, respectively, and the hysteresis evolution of the dynamics during decompression.

The ever-slowing dynamics at 6.3 GPa strongly resembles the physical aging usually observed in thermally activated structural relaxations, associated to the interplay between density changes and MRO ordering processes [24–26]. In this regime, the corresponding ISFs evaluated at successively larger waiting times, $t_w$, elapsed from the pressure change, shift continuously toward longer decay times (Fig. 4a) and can be rescaled into a master curve when normalizing δt by τ (Fig. 4b). The validity of the temporal scaling confirms the homogeneous nature of the collective motion. The corresponding evolution of τ as a function of $t_w$ is reported in the inset of Fig. 4b) and echoes the results obtained in MGs at atmospheric pressure and high temperature [26], that is a first rapid aging regime which obeys a phenomenological equation $\tau(t_w) = \tau_0 \exp(t_w/\tau^*)$ followed by a constant dynamical state (last point at $t_w$>8000s, excluded from fit), less visible here. The yellow line in the inset corresponds to a fit of the previous equation to τ($t_w$). It yields $\tau^*$=2300s and $\tau_0$=34s, respectively compatible to and ten times smaller than atmospheric pressure high temperature literature data [25–27]. This means that despite the rejuvenation at early stages after pressure compression (and therefore the slower value of $\tau_0$), the rate of aging is similar in both temperature and pressure studies.

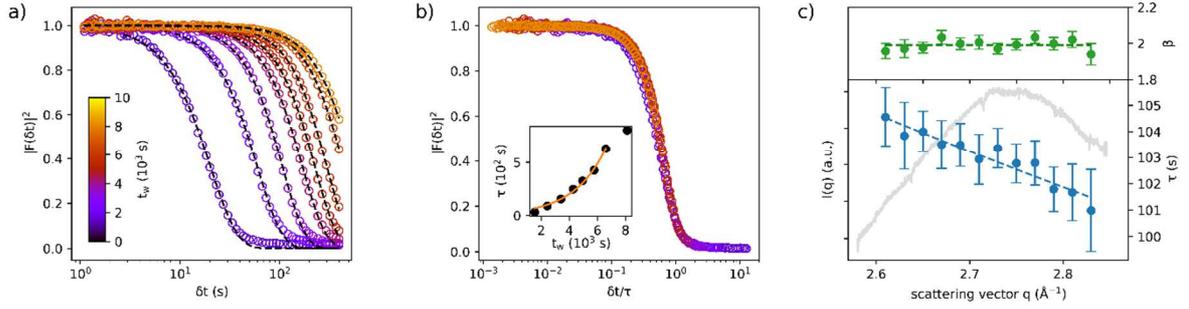

**Fig. 4 | Aging and wave-vector dependence of the dynamics in the homogeneous regime**. a) ISFs measured at 6.3 GPa as a function of the elapsed time from pressure change. Black dotted lines correspond to KWW fits to the data. Aging is visible through the shift to long delay times with increasing waiting time (from left to right ISFs). b) Scaling of the ISF as a function of the reduced time $\delta t/\tau$. Inset shows the evolution of the corresponding $\tau$ and the best fit to the equation $\tau(t_w) = \tau_0 exp(t_w/\tau^*)$ (line). c) Wave-vector dependence of the dynamics at 6.3 GPa: top) KWW shape parameter; and bottom) relaxation time (symbols) and integrated intensity in the detector (grey line) measured at 6.3 GPa.

Interestingly, the same physical mechanism seems to control the atomic rearrangements in both the rejuvenation and relaxation regimes. All data can be described by compressed ISFs with an averaged compressed shape parameter, β, ranging from 1.6 to 2, depending on the degree of heterogeneity of the dynamics. Similar compressed values of β have been reported in all MGs under temperature studies [26,28,29]. Thanks to the high signal to noise ratio of the data and the large area detector used during the XPCS measurements, we can evaluate the dynamics of the glass at different wave-vectors $q$ even in the nonergodic state, bypassing the problem of aging [27]. Although the probed q-range is limited by the size of the detector (Fig. S6), the relaxation time follows a $\tau(q)=1/cq^\alpha$ dependence from the probed wave-vector, with 0<α<1. This is shown in Fig. 4c for 6.3 GPa where the fit yields α=0.36±0.04. The wave-vector dependence of the dynamics and the constant compressed shape of the ISFs implies that the ISFs can be described by $|F(\delta t)| = e^{-(\delta t/\tau(q))^\beta} = e^{-(c^\theta \delta t^\theta q)^k}$ with θ=1/α=2.74 and k=α·β= 0.73 at 6.3 GPa [30]. This expression confirms the complex nature of the dynamics of MGs and contrast with the high temperature diffusive motion of liquid metals which would instead corresponds to τ(q)=1/Dq² and thus to $|F(\delta t)| = e^{-Dq^2 \delta t}$, with D the diffusion coefficient.

To characterize the evolution of the dynamics over the complete compression/decompression cycle, we have defined a dynamical heterogeneity parameter, in the following way. We first extract the temporal evolution of the correlations $C(t, \delta t = \tau) = \langle I(t) \cdot I(t+\tau)\rangle/\langle I\rangle^2$ at a fixed delay time between frames corresponding to the structural relaxation time τ obtained by the KWW analysis of the individual ISFs (red curve in Fig. 5a). We then compute distributions of the correlation values observed at a single pressure (Fig. 5b) by averaging all the different histograms of $C(t, \delta t = \tau)$ at this pressure. Heterogeneous dynamics lead to broad, potentially multimodal dynamics as illustrated by the distribution obtained at 3.3 GPa, where two main distinct contributions are visible in addition to the long tail at large values. Overall, the distributions broaden toward both the low and high correlation values when pressure increases from 0.1 GPa to 3.3 GPa, and shrink afterward. As the width of the distribution translates directly to the behaviour of the dynamics, we define a heterogeneity parameter ΔC as the smallest width that contains 90% of the values of the statistics of the distribution (as shown in Fig. 5b at 0.1GPa). Similar results are observed also for lower percentages of ΔC (Fig. S9). The evolution of this heterogeneity parameter shows the pathway of the dynamics during the compression-decompression cycle, and is displayed in Fig. 5c). The compression regime is inversely related to the evolution of $\langle\tau\rangle$, with the bell shaped curve centred around 3.3 GPa, and corresponds to the dynamical transition between the rejuvenated and relaxed regimes described above. Interestingly, the decompression pathway shows the hysteresis deduced from the TTCFs at 3.3 GPa for increasing and decreasing pressures (Fig. 3). The decreasing pressure does not impact significantly the dynamical behaviour until 0.5 GPa, with a heterogeneity that remain relatively constant, possibly going through a limited increase. At 0.5 GPa, the heterogeneity rises up to a value similar to the maximum observed in compression. This pressure step corresponds to a fully deflated membrane in the DAC, and one could associate the dynamic fluctuations to mechanical instabilities of the cell. The pressure stability of 0.04 GPa over the course of the measurement dismisses however this possible artefact.

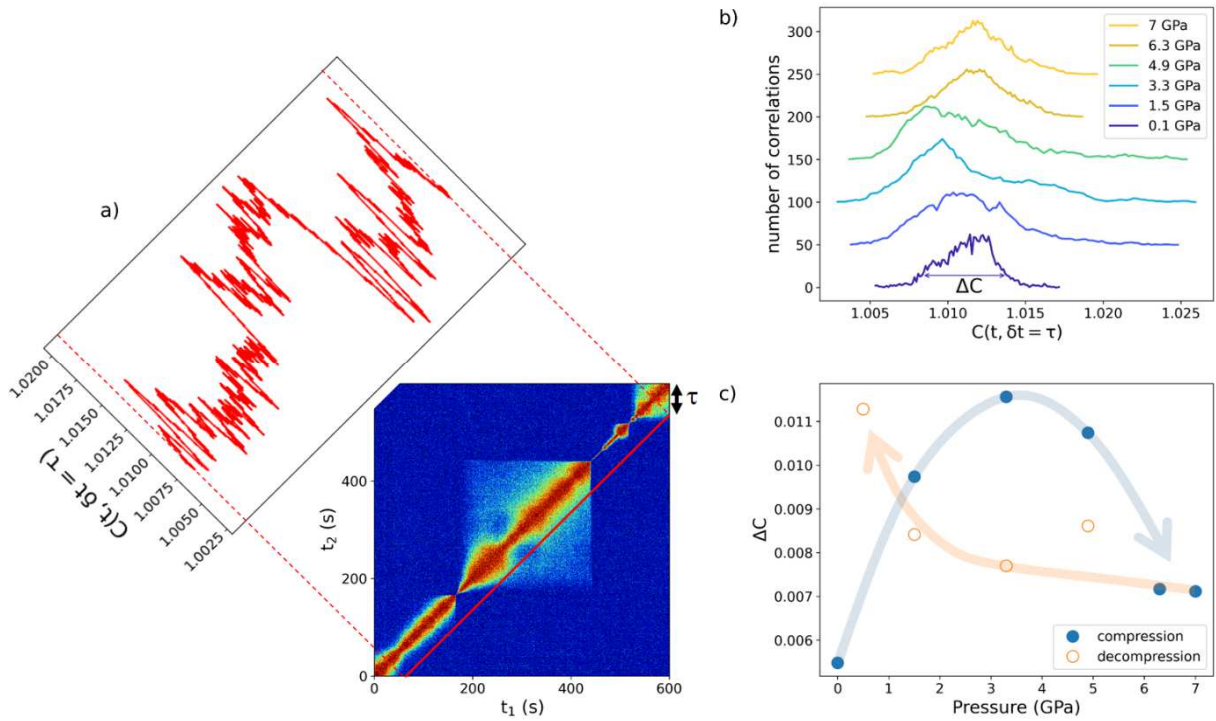

**Fig. 5 | Dynamical pathway during full compression-decompression cycle.** a) Typical evolution of $C(t_1, t_2) = \langle I(t_1) \cdot I(t_2)\rangle/\langle I\rangle^2$ at a fixed delay time $\delta t = t_1 - t_2 = \tau$. b) Corresponding distributions of $C(t, \delta t = \tau)$ during the compression stage for $\tau$ the structural relaxation time obtained from the KWW fits of the ISFs. Distributions are offset vertically for clarity. c) Dynamical heterogeneity ΔC as a function of the applied pressure. This parameter represents the width of the distributions in panel b), defined as the smallest interval that contains 90% of the correlations. The first point corresponds to the loading pressure of 0.1 GPa. The fixed delay time $\delta t = \tau$ is not accessible at 1 atm because dynamics is too slow to observe a full decorrelation in the TTCFs.

**Discussion**

The existence of two dynamical regimes controlling the atomic motion of MGs under hydrostatic pressure is consistent with results from recent theoretical works, which suggest that increasing pressure leads to the formation of a second metastable higher-energy state in the potential energy landscape [16,17]. In this picture, fast dynamics correspond to temperature-assisted transitions within this two-level system which leads to rejuvenation at low pressures. With further increase of pressure the second metastable state vanishes, and dynamics reverse then to the slow structural relaxation, similarly to our data [16,17]. It would be interesting to know, whether the model could describe also the heterogeneous to homogeneous evolution of the particle motion during compression.

The decorrelation events observed in the rejuvenation regime, especially when fast and complete decorrelation occurs, are the sign of cascade or avalanche-like cooperative relaxation mechanisms, where local relaxation events trigger neighbouring events in a chain reaction[31]. While the trigger for thermally activated relaxation in MGs is highly localized and independent of the stability of the system [32,33], this chain reaction implies a high spatial density of local minima in the PEL of the glass [34], as isolated minima do not interact with each other. Such avalanche-like dynamics have been reported as an aging mechanism in the similar $Pd_{43}Cu_{27}Ni_{10}P_{20}$ metallic glass [29], in a mechanically stressed metallic glass ribbon [28], and as a mediator of aging and/or crystallization in a hard-sphere glass [33,35]. Regardless of the final structural state (aged glass or crystal), Yanagishima et al. showed that the avalanche events statistically appear in regions of lower local density and bond orientational order [33], reinforcing the heterogeneity of the PEL mentioned above at low pressures. Therefore, the avalanches-like dynamics observed at low pressures witnesses a higher degree of inhomogeneity in the glass structure in this pressure range, in agreement with the as-cast nature of our glass. As individual avalanches do not necessarily increase the local order in the glass [33], and longer time is necessary for the aging trend to emerge at low pressures, the rejuvenation regime persists for several steps in pressure and for many hours per pressure without any signature of relaxation. The transition from rejuvenation to aging hints also toward an effect of the excess free volume, which is present in the as-cast glass but seems to reduce greatly during the relaxation at high pressures, as suggested by the dynamical hysteresis, even if further measurements would be necessary to investigate more this aspect. XRD studies report the occurrence of elastic deformations during the compression of MGs, supporting the idea of a homogeneous fractal network model for the glass [36], as opposed to the heterogeneous structural model of liquid-like regions of loosely bonded atoms embedded in a solid-like matrix [3,37,38]. Our work shows that the presence of apparently reversible structural changes under hydrostatic pressure compression (Fig. 2b), is not a sufficient condition to assure a simple elastic structural mechanism under compression, as they can be accompanied by a dramatic hysteresis evolution of the dynamics (Fig. 3 and S5).

The $\tau(q) \approx 1/q^\alpha$ wave-vector dependence of the relaxation time implies a super-diffusive collective particle motion in the glass at all pressures, which differs from the well-known structural dependence of the relaxation time observed in supercooled liquids in the proximity of the FSDP [39,40]. Above the glass transition temperature, the equilibrium dynamics is associated to cage-escape processes, and the long-time collective motion is sub-diffusive leading to a stretched exponential decay of the ISFs, described thus by a value of $\beta<1$ [40,41]. In the glassy state, atomic mobility of MGs originates from fast secondary relaxation processes, such as the β- and γ-processes [3,42]. These processes control the stress response of the material in the non-ergodic state and have been associated to cooperative string-like particle motions in nanometric liquid-like regions [43,44]. Compressed ISFs and super-diffusive dynamics have been reported in many different complex systems as colloidal gels, clays, concentrated emulsions, oxides and soft colloids [30,45,46]. In these systems, the anomalous dynamic has been associated to the presence of random local stresses in the materials, which are then released triggering the faster-than-exponential collective dynamics [30,31,46–48]. In MGs this stress propagation could be related to the kinetics of structural rearrangements induced by the stress field controlled by the β- and γ- relaxation processes. Further studies will be necessary to clarify the nature of the collective dynamics in MGs and their evolution paths under annealing and pressure.

**Methods:**

Glass synthesis: We prepared a PtCuNi precursor by arc-melting the pure metallic components (with purity >99.95%) under a Ti-gettered Ar-atmosphere (with purity >99.999%). We then alloyed inductively the elemental P with the PtCuNi precursor in a fused-silica tube under Ar-atmosphere. In order to obtain as low as possible oxide content, the alloy was subjected to a fluxing treatment in dehydrated $B_2O_3$ for more than 6 hours at 1473 K. The ribbons were produced by melt spinning of the master alloy on a rotating copper wheel under high-purity Ar atmosphere. The resulting glass ribbons of $Pt_{42.5}Cu_{27}Ni_{9.5}P_{21}$ at.% had a thickness of 20 µm and a width of 2 mm.

High Pressure: the sample was cut from the as-cast 20 µm thick ribbon to a rough shape of 50x50x20 µm³. The sample was subsequently pre-loaded at 0.1 GPa in a membrane driven Diamond Anvil Cell with a ruby sphere and 4:1 methanol/ethanol mixture as pressure-transmitting medium (PTM), to ensure a perfectly hydrostatic compression up to 10 GPa [49]. The DAC was equipped with 600 µm diamonds (culet size) and a pre-indented laser drilled stainless steel gasket to make a 60 µm x 300 µm (height x diameter) experimental volume. The compression cycle up to 7 GPa is shown in Fig. S1: similar pressures were reached in compression and decompression, and the elapsed time at each pressure was around three hours in compression and one hour in decompression. The pressure was measured from the wavelength of the Chromium $^2E \rightarrow {}^4A_2$ transition in a ruby sphere after and before each pressure change, and a dedicated pressure protocol on the membrane ensured a pressure variation lower than 0.12 GPa at all pressures (Fig. S7).

X-Ray Diffraction: The structure of the metallic glass under pressure was monitored by two different runs of x-ray diffraction. The first run, conducted at beamline ID27 at ESRF synchrotron, France, reproduced the pressure protocol of the XPCS experiment. Experiment was performed using an incident energy of 33 keV, an EIGER2 X CdTe 9M (active area = 233.1 x 244.7 mm², pixel size = 75 µm) detector and a DAC loaded with 4:1 methanol/ethanol mixture as PTM, a sample and a ruby sphere for pressure determination. Background was collected at each pressure by measuring the scattering pattern of a location inside the DAC next to the sample. The maximum scattering vector probed in this run is q=12 Å⁻¹. Azimuthal integration of the 2D scattered patterns was performed using the pyFAI python library [50,51] to yield 1d diffraction patterns, and the computation of the (background corrected) Faber-Zimman structure factor with Krogh-Moe-Norman normalization [52] was performed using the python-based Amorpheus software [53].

To assess quantitatively the link between the peak position and the sample density, x-ray diffraction data was collected as a function of temperature at atmospheric pressure to compare the shift of the first sharp diffraction to the coefficient of thermal expansion measured by dilatometry (Fig. S1). The XRD data was collected at the beamline ID15a [54] at the ESRF synchrotron, France. Data acquisition using an incident beam energy of 68.5 keV and the scattered diffraction pattern was collected with a Pilatus3 X CdTe 2M detector (active area = 253.7 x 288.8 mm², pixel size = 172 µm). A sample to detector distance of 1.087m was chosen to maximize the resolution on the first sharp diffraction peak. The background was acquired in the same condition with an empty sample. Diffraction patterns were azimuthally integrated using routines from the pyFAI library [50,51], and locally implemented corrections for outliers rejection, background, polarization of the X-rays and detector geometry, response, and transparency, to yield 1D diffraction patterns.

XPCS: In order to optimize high-energy and high-pressure XPCS studies, we performed three different XPCS campaigns for a total of 3 weeks of beamtime at beamline ID10 at the ESRF synchrotron, France. The main data have been collected by using a 20.95 keV partially coherent monochromatic beam with a photon flux of 4.2x10$^{11}$ photon/s, focalized by a 2D Be lens transfocator to 50.5x14.2 μm$^2$ (HxV, FWHM) cut by a pair of slits for an illumination area of 8x8 μm$^2$ on sample. The second sample was measured in a second run with an incident energy set to 21.67 keV with a flux of 7.3x10$^{11}$ photon/s focalized to a beamsize of 5.2x4.2 μm$^2$ (HxV, FWHM) on sample. To record the speckle patterns, we placed an Eiger2 4M CdTe detector (active area = 155.1 x 162.2 mm$^2$, pixel size = 75 μm) 5 meters downstream at an angle corresponding to the pressure-dependent position of the FSDP, whose maximum is at 2.79 Å$^{-1}$ at atmospheric pressure and 25°C. The top part of this FSDP is reconstructed by integrating the intensity in the detector, allowing the monitoring of the position of the peak during the measurements. An additional PILATUS detector has been also employed to control the evolution of the structure in a broader Q range during the experiment. A constant acquisition time of 0.1s/frame was kept throughout the whole XPCS experiment, with scans ranging from 6000 frames to 14000 frames depending on τ. Intensity-Intensity correlation functions, g$_2$(t), and TTCFs are extracted from the successive speckle patterns using the event correlator method described in [55]. The ISFs are then obtained from the g$_2$(t) through the Siegert relation $g_2(q, \delta t) = 1 + \gamma \cdot |F(q, \delta t)|^2$, whose validity in non-ergodic systems is assured by the use of large area detectors [56,57]. In this expression γ is the experimental contrast related to the degree of coherence of the beam. TTCFs have been evaluated from the normalized correlation $\langle I(t_1) \cdot I(t_2) \rangle / \langle I \rangle^2$ between all pairs of scattering patterns recorded during a scan at a given q. The main diagonal corresponds to the elapsed time of the measurement with *t(frame 1) = t(frame 2) = t*, while any point off this diagonal express the correlation value at a certain delay time *δt = t(frame 2) – t(frame 1)* after the first frame is recorded. To quantify the evolution of the dynamics with pressure, we extracted the characteristic times τ of all scans acquired during the compression by fitting KWW functions to the F(q,δt) data. We further averaged the different values of τ at a single pressure, to get an average ⟨τ⟩ for each isobars. No contribution to the dynamics has been observed from the background (diamonds and pressure transmitting medium) as shown in Fig. S8. For the analysis of Fig. 5, the computation of $\langle I(t) \cdot I(t+\tau) \rangle / \langle I \rangle^2$ have been done on the raw data, i.e. taking into account also the aging within each scan. Although this potentially leads to an overestimation of the heterogeneity parameter in the aging regime, we found this effect to be very limited leading to a well-defined transition between the rejuvenation and aging regimes.

**Acknowledgements**

We acknowledge ESRF (Grenoble, France), for the provision of experimental facilities. Parts of this research were carried out at ID10 and ID27 beamlines under the LTP project HC4529. We gratefully thank M. di Michiel for providing in-house experimental time at the ID15a beamline and for his assistance during the experiment. We would also like to thank T. Poreba, K. Lhoste and D. Duran for assistance. This project has received funding from the European Research Council (ERC) under the European Union's Horizon 2020 research and innovation programme (Grant Agreement No 948780). All data needed to evaluate the conclusions in the paper are present in the paper and/or the Supplementary Materials.

**Competing Interests**

The authors declare that there are no financial or non-financial competing interests.

**Author Contributions**

B. R., A. C., Y. C. and F. Z. conceived the study. N.N. and M.F. prepared the samples. G. G., J. J. and M. M. provided technical and scientific support for all high pressure experiments. A. C., B. R., F.Z., Y. C., S. L., T. D., N. N., M. F., E. P., J. S. and A.R. conducted the HP-XPCS experiments at beamline ID10. A.C., B.R., S. L., G.V. and M. di M. conducted the high temperature XRD measurements at beamline ID15A. A.C., J.S., B.R. and G.G. performed the HP-XRD experiments at beamline ID27. A.C. analysed all data with the support of B.R., Y.C., G.V., G.G. and G.M..  A.C. and B.R. wrote the manuscript with inputs from all authors.

# Supplementary Materials to "Denser glasses relax faster: a competition between rejuvenation and aging during *in-situ* high pressure compression at the atomic scale"

*A. Cornet et al.*

1. <u>**Compression decompression cycle and First Sharp Diffraction Peak evolution:**</u>

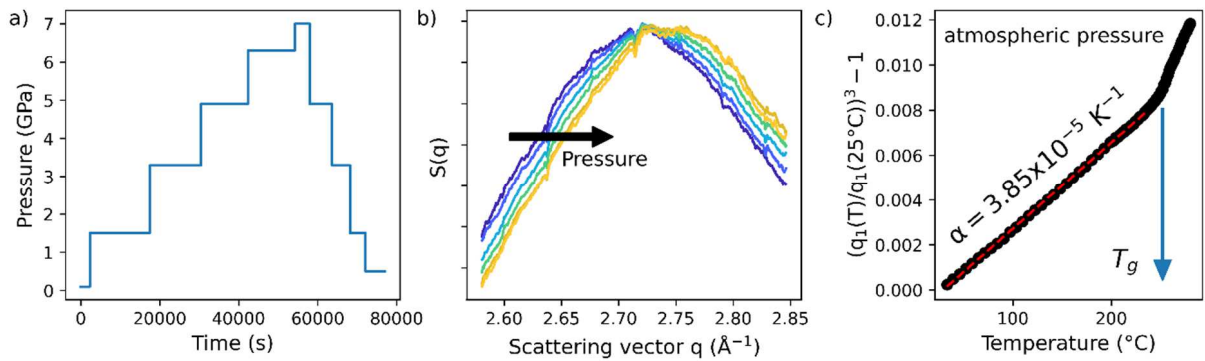

**Figure S1** – **a)** Pressure protocol. All XPCS measurements are performed during the different isobars. **b)** Evolution with pressure of the top of the diffraction peak during the XPCS measurements. Data are reconstructed by integrating the detector intensity. For clarity, the curves correspond to the compression stage only. **c)** Evaluation of thermal expansion coefficient from the evolution of the FSDP position $q_1$ with temperature measured at atmospheric pressure with synchrotron high energy XRD.

Similar pressures were reached upon compression and decompression to allow direct comparison. The first step in compression at 0.1 GPa corresponds to the preloading of the cell, and the pressure of the subsequent steps are of 1.5, 3.3, 4.9, 6.3, and 7 GPa. The last decompression step at 0.5 GPa corresponds to the situation where the diamond anvil cell (DAC) membrane is fully deflated, but the DAC remains mechanically locked. The top of the First Sharp Diffraction Peak (FSDP) can be reconstructed by integrating the intensity collected on the detector during a complete XPCS scan. The quantitative estimate of the peak position $q_1$ is used to verify the consistency of the XPCS experiment with X-ray diffraction (XRD) experiment. Here, the continuous shift of the FSDP toward the high scattering vectors confirms the XRD results shown in the main text. $q_1$ being linked to the characteristic distance $\ell$ of the medium range order of the glass by $q_1 = 2\pi/\ell$, we can quantitatively assess the validity of the link between q1 and the macroscopic density of the glass by comparing the coefficient of thermal expansion (CTE) derived from high energy XRD measurement to the CTE obtained from dilatometry. The CTE $\alpha$ is inferred from the evolution of $(\rho(T)-\rho(25°C))/\rho(25°C) \propto (q_1(25°C)/q_1(T))^3 - 1$. We obtain $\alpha = 3.85 \times 10^{-5}$ K$^{-1}$ in the glassy state, in good agreement with the reported value obtained by dilatometry[1] $\alpha = 3.95 \times 10^{-5}$ K$^{-1}$. This shows that the FSDP is directly linked through the macroscopic density for this glass.

## 2. Ruling out artefacts as the source of the heterogeneous dynamics

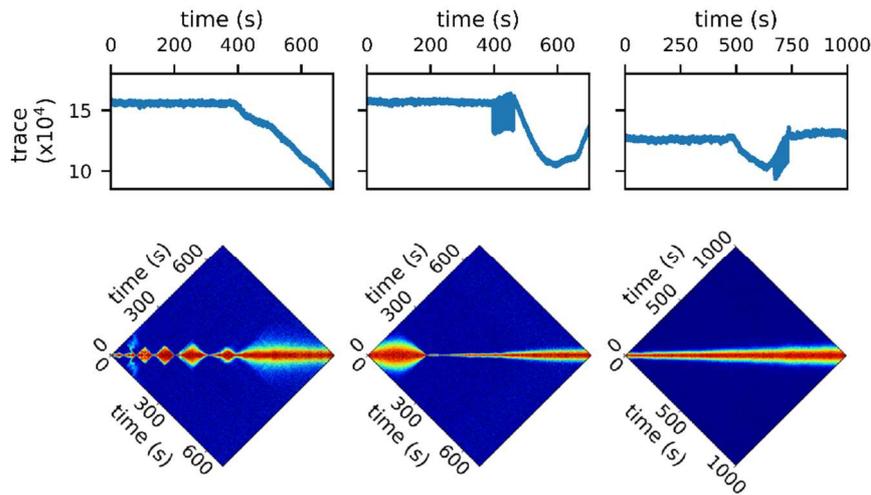

**Figure 6** - Trace (total intensity on detector) and selected TTCFs in compression at 3.3 GPa (left panels, elapsed time = 9700s), 4.9 GPa (central panels, elapsed time = 9800s) and in decompression at 3.3 GPa (right panels, elapsed time = 860s).

Although the intensity impinging on the sample is in general stable, some fluctuations can occur during a full week of beamtime, and are usually due to adjustment of the electron beam in the storage ring and refilling mode. As the magnitude of these fluctuations is usually small compared to the total intensity, data are not affected. In Figure S2 we report selected TTCFs and the corresponding trace (total intensity in the detector as a function of time) which show two different situations. In the left panel and the beginning of the central panel, heterogeneous dynamics appear while trace is stable. Differently, in the central and right panel trace shows fluctuations related to the re-fill in the storage ring with no influence on the dynamics. This demonstrates that fluctuations of the incoming beam intensity are not responsible for the observation of heterogeneous dynamics. These data allow us to rule out also possible sample movements as sources of induced decorrelations. The position of the sample was monitored by a microscope before and after each scan. Large movements on the scale of 10 µm (>5 times the Rayleigh Criterion) would then been detected, which is not the case. We exclude also smaller, micrometres movements as if the decorrelation would be associated only to a change of the scattering volume, the decorrelation time τ should be identical before and after the event, which is generally not the case. In both the left and central panel, the 'decorrelations' in the TTCFs lead to different dynamic profiles, which implies that the relaxation of the systems has changed. Another example is reported also in Fig. S3 as described below.

It is well-known that some Pressure Transmitting Medium (PTM) can alter the property of a glass under compression. This is the case for instance of gas loading with He that can enter the large open network of silica glasses[2]. Due to its large molecular structure this is not the case for the alcohol mixture chosen here as PTM even in the presence of large open structures[2].

## 3. Additional data on the rejuvenation and relaxation regimes

Fig. S3 shows the TTCFs measured at 4.9 GPa in compression. The data shows aging regimes separated by a cascade relaxation at 9600s. The mix of cascade relaxation and aging between the two well defined dynamical regimes at 3.3 and 6.3 GPa shows the transition is not abrupt but continuous. The third TTCF at 4.9 GPa is also a further confirmation of the absence of sample movement as the source of the heterogeneous dynamical regime observed at low pressures. If sample movement caused the decorrelation event at 9600s, decorrelation time τ should be identical before and after the event, which is obviously not the case. The TTCFs at 0.5 GPa at the end of the decompression stage show the heterogeneous dynamical regime is eventually recovered but at a lower pressure compared to compression, in accordance with the evolution of the dynamical heterogeneity introduced in the figure 5.

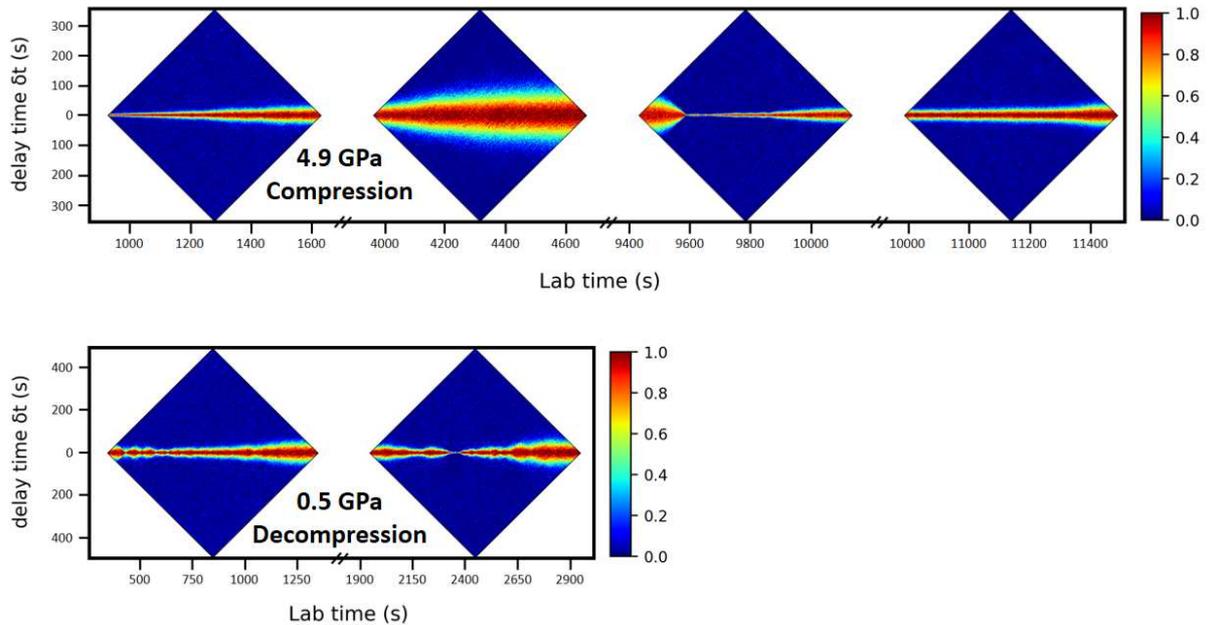

**Figure S3** - TTCFs obtained at 4.9 GPa during compression (top row) and 0.5 GPa in decompression (bottom row).

### 4. Repeatability: results from a second experiments

We controlled the repeatability of the results on a new sample measured in a different run. As shown in Fig. 2e and Fig. S4, results of both runs overlap and show the same heterogenous vs homogeneous transition with pressure, which confirm the robustness of the data showed in this study.

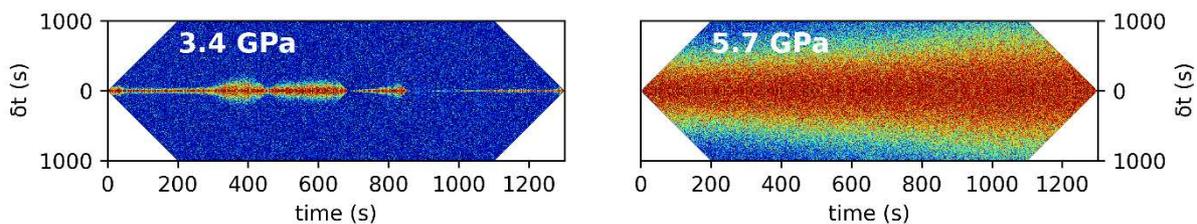

**Figure S4:** Partial TTCFs measured in a second sample in the low pressure heterogeneous (left) and high pressure homogeneous (right) regimes.

### 5. Hysteresis evolution of the dynamics under pressure compression and decompression

The hysteresis shown in the main text from the comparison of the TTCFs at 3.3 GPa and the complete pathway of the heterogeneity parameter is also visible from the characteristic relaxation time of the intermediate scattering function. In the figure S5 we represent intermediate scattering functions (ISFs) from the compression and decompression stages at 1.5 GPa and 3.3 GPa. To provide an accurate comparison, we chose to plot the curves obtained at the most similar elapsed times $t_w$ (defining the duration of the isobar). Both the shift of the curves and the relaxation times inferred from the KWW model fitted to the data show slower dynamics during the decompression stage, confirming the hysteretic behaviour.

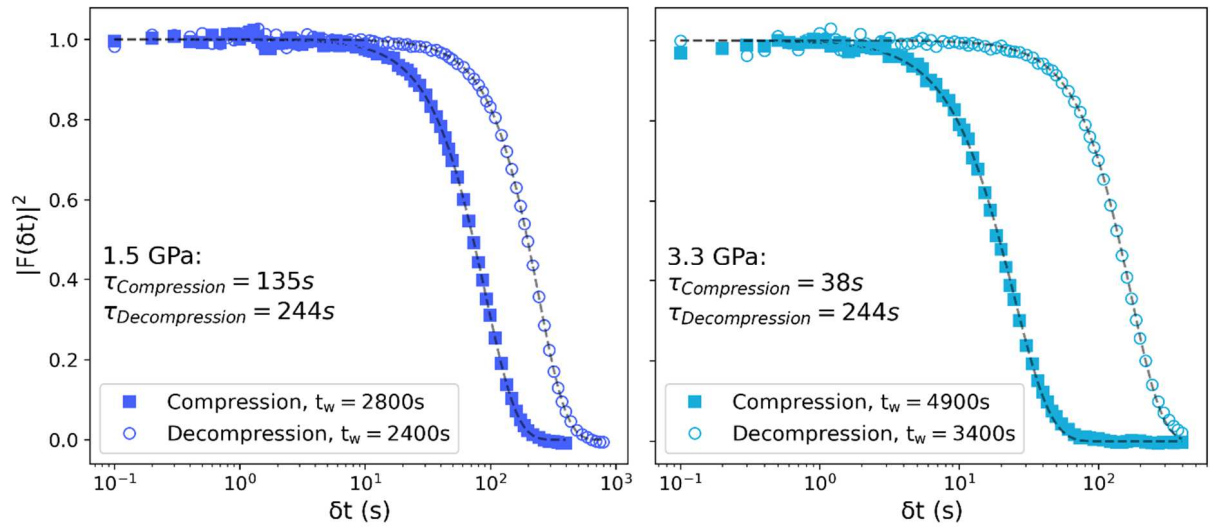

**Figure S5:** ISFs measured at 1.5 GPa and 3.3 GPa in compression and decompression at similar elapsed times $t_w$, showing the hysteresis within a pressure cycle. Dashed lines correspond to fits of the KKW model to the data. The corresponding relaxation times τ are reported.

### 6. Wave-vector dependent XPCS study

To determine the dependence of the ISF with respect to the scattering vector q, we added a binning on the raw data. In the figure S6 we plot a colour representation the total scattered intensity in the detector within a single scan of 7000 frames with an acquisition time of 0.1s. The distribution of the intensity clearly shows the maximum of the diffraction peak. The grey area corresponds to the raw mask applied during the pre-processing of the data, which covers the shadow of the vacuum tube between the sample and the detector and Kossel lines from the diamonds. On the right panel, the segmentation of the unmasked area of the detector into several bands at different is visible. The TTCFs and ISFs were then extracted for the data corresponding to each of these bins.

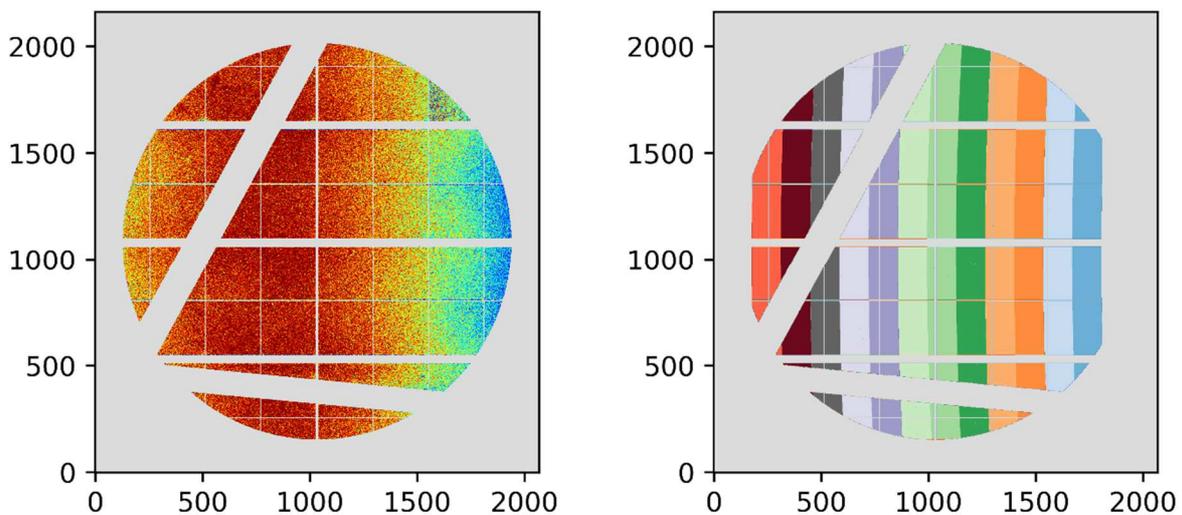

**Figure S6:** Left panel: Integrated intensity in the detector during an XPCS scan. A portion of the reddish ring associated to the FSDP is well visible. Right panel: Q-binning of the data to extract the q-dependence evolution of the ISFs.

## 7. Pressure protocol and stability

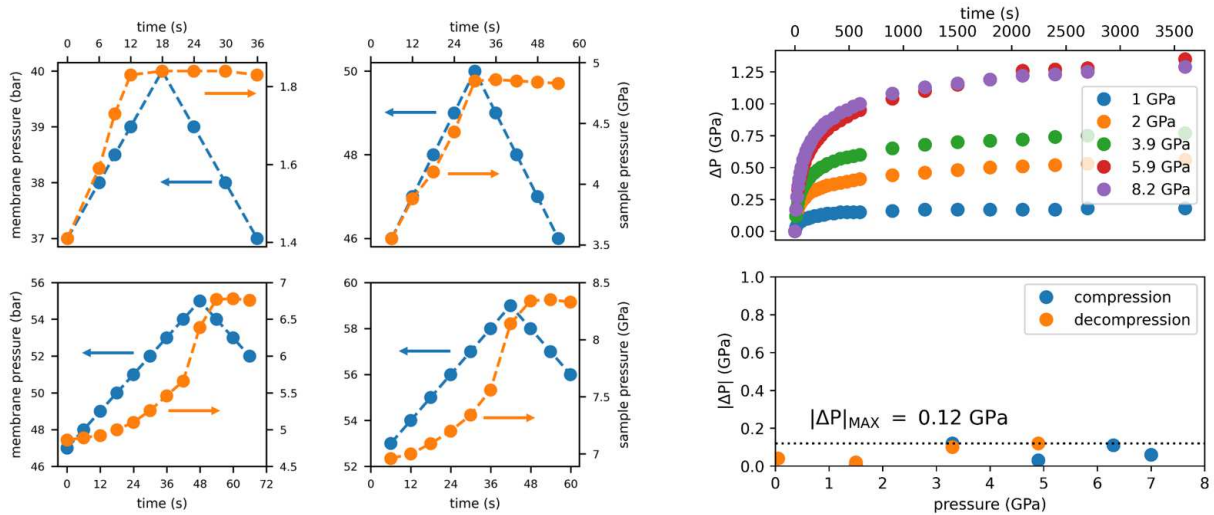

**Figure S7** - Loading protocol for pressure stability. Simultaneous recording of the membrane and ruby pressures (left panel) showing the stabilisation effect of the loading protocol. Pressure overshot *without* the adapted pressure protocol (upper right panel) and pressure drifts recorded during the experiment using the adapted protocol.

As XPCS is extremely sensitive to any structural change, it is essential to minimize any potential pressure drift during the measurements. The typical pressure increase after reaching the desired set-point for our DAC configuration is shown in the upper right panel, and can be higher than 1 GPa over one hour. To mitigate this issue, we have determined how much we can reverse the membrane pressure to stabilize the pressure on the sample. More precisely, we measured how much one can decrease the membrane pressure after an initial increase before we can see any change in the sample pressure (down to our precision of 0.01 GPa), as shown in the left panels. We have applied this loading protocol during the measurements, and we report all pressure drifts, taken as the pressure difference between the beginning and the end of a pressure steps. We can see the pressure drifts are now limited to about 0.1 GPa. Importantly, this drift is random and does not depend on the nominal pressure, so it is not responsible for the two-steps pressure effect on the dynamics reported in this study.

## 8. Diamond Anvil Cell XPCS background

Fig. S8 contains two ISFs obtained under pressure with beam focused on the sample or in the experimental volume next to the sample. The absence of a decorrelation in the second case demonstrates that the contribution of the Diamond Anvil Cell, which comprises contributions from the diamonds and the pressure-transmitting medium, are only static contributions. The sample dynamics probed with XPCS is therefore not affected by the high pressure cell.

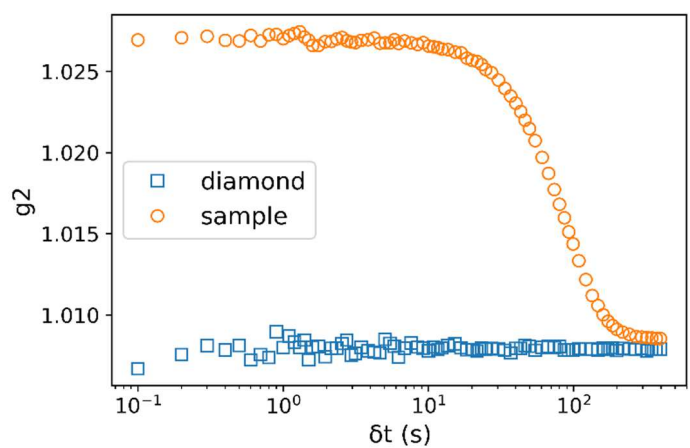

**Figure S8** - Correlation function g2 from beam targeting the sample (orange) and beam out of the sample, showing only the contribution of the diamond and pressure transmitting medium.

## 9. Dynamical heterogeneity parameter

The dynamical heterogeneity parameter ΔC characterize the level of inhomogeneity of the glass atomic scale dynamics. This parameter corresponds to the smallest width that encompasses 90% of the distribution of the correlation values at a fixed delay time $\tau$ for each pressure. This 90% threshold was chosen to reflect the extremes values taken by the correlation values. However, we show that the evolution of with respect to pressure is not threshold strictly dependent on the value of this threshold. In the figure S9 we reproduce the figure 5c) of the main text for different threshold values: 90%, 80%, 70%, and 60% (upper left, upper right, lower left and lower right panels respectively).

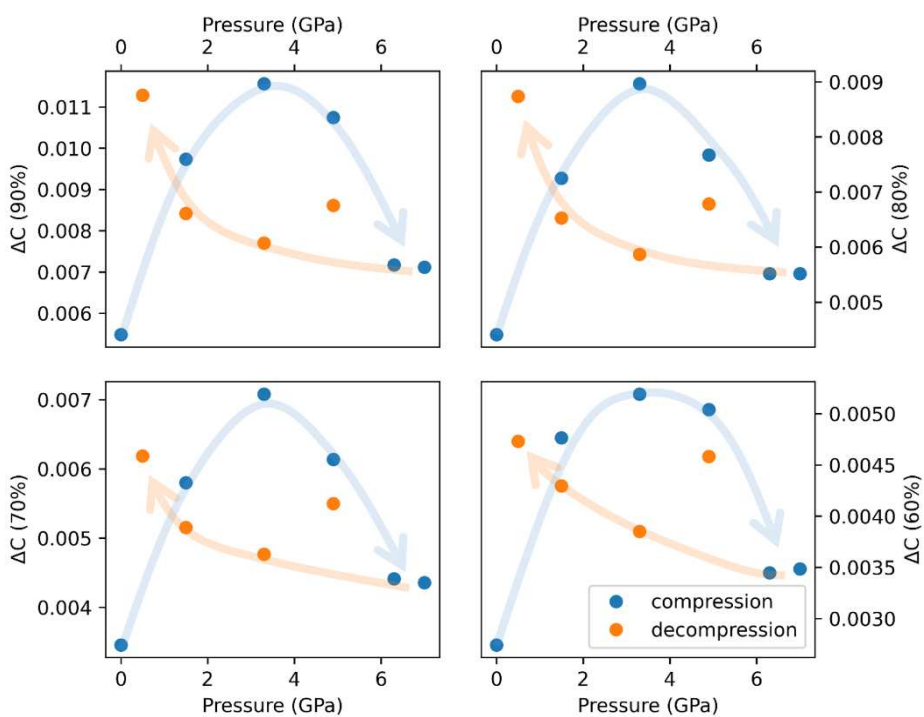

**Figure S9** – Dynamical heterogeneity parameter as a function of pressure for different threshold value: 90% (top left), 80% (top right), 70% (bottom left) and 60% (bottom right).

The result obtained for a width of 90% is reproducible quantatively down to a width of 70%, and qualitatively down to a width of 60%. Overall, this confirms the robustness of the heterogeneity parameter ΔC.